\documentclass[12pt]{iopart}

\usepackage{iopams}  
\usepackage{graphicx}

\begin{document}

\title[A Multi-Layer Three Dimensional Superconducting Nanowire Photon Detector]{A Multi-Layer Three Dimensional Superconducting Nanowire Photon Detector}
\author{A. Matthew Smith}
\address{Air Force Research Lab, Information Directorate, 525 Brooks Rd.,  Rome, New York 13441, USA}
\ead{amos.matthew.smith.ctr@rl.af.mil}
\begin{abstract}
Here we propose a new design paradigm for a superconducting nanowire single photon detector that uses a multi-layer architecture that places the electric leads beneath the nanowires.  This allows for a very large number of detector elements, which we will call pixels in analogy to a conventional CCD camera, to be placed in close proximity.  This leads to significantly better photon number resolution than current single and multi-nanowire meanders, while maintaining similar detection areas.  We discuss the reset time of the pixels and how the design can be modified to avoid the latching failure seen in extremely short superconducting nanowires.  These advantages give a multi-layer superconducting number-resolving photon detector significant advantages over the current design paradigm of long superconducting nanowire meanders.  Such advantages are desirable in a wide array of photonics applications.
\end{abstract}

\pacs{85.25.Oj,74.78.-w}
\maketitle

\section{Introduction}
Construction of photo-counting devices with both high counting efficiency, high number resolution and short reset times, is highly desirable for a wide array of application in quantum key distribution (QKD) \cite{xu}, quantum communication \cite{obrien}, quantum computing \cite{klm,uskov1,knill} among others \cite{hadfeild,smith1}.  Here we describe and perform some simple analysis of a proposed detector design that uses multiple short sections of superconducting nanowires.  We refer to these short sections of nanowire as pixels and arrange them in a two dimensional grid in analogy with a standard CCD camera.  We will discuss the potential advantages in such a system.  

We will only briefly describe the operation of superconducting nanowires.  When an incident photon strikes a Niobium nitride (NbN) nanowire, or other superconducting material such ad NbTiN or a-W$_x$Si$_{1-x}$ developed recently at NIST, it creates a resistive hotspot \cite{NIST}.  This hot spot causes the current in the superconductor to deflect around the spot, thus increasing the current density in the wire.  This increased current destiny leads to an increase in the temperature of a small section of the wire.  If the nanowire is held just bellow the critical current for superconduction, then the increase in heat will break the superconducting condition and the resistance of the wire will spike upward for a short time.  This resistance spike creates a measurable current in the external resistance load and a photon is counted.  For specific detail see \cite{NIST,gurevich,dauler} among others.

     Present superconducting nano wire systems, such as NbN meanders, have reasonably good counting efficiency \cite{dauler,marsili}, by which we mean the probability of an incident photon being detected at over $25\%$.  However, a significant problem exists with number resolution and relaxation time and fill factors\cite{gurevich,dauler,marsili}.  A detector consisting of a single wire can be made to cover a significant detection area by creating a meander.  Usually this means folding the nanowire back and forth across the desired area of approximately $10\mu$m x $10\mu$m \cite{dauler,marsili}.  This however is not a number resolving detector.  All that the detector can feel is the loss of the superconducting condition of the nanowire.  Should two photons strike the wire simultaneously in two different locations the current drop is very similar.  One suggested solution to this lack of number resolving capability is to increase the number of wires in the meander.  This has been done experimentally by Dauler {\it et. al.}\cite{dauler}.  While this approach improves on the single wire meander it still consists of long wires each of which occupies a significant potion of the active detection area, (i.e. each wire in a 4 wire meander takes approximately $25\%$ of the active area).  In order to have a high probability of correctly detecting $n$ number of photons one would need significantly more than $n$ wires.
     
     A second drawback of the long nanowire approach is the relaxation time of the detector.  It has been shown that the relaxation time, the time for the hot spot to dissipate, is related to the kinetic-inductance of the nanowire \cite{kerman}.  
This leads to a relaxation time of about 10ns for a $10\mu$m x $10\mu$m meander.  The operational repetition rate, for a non-deterministic operation, will need to be slower than the relaxation time, to avoid interactions between the relaxation and incoming photons.  This leads to rep. rates much slower than current pulsed laser systems which are capable of GHZ frequencies.
     
     We now propose a detector design, patent pending, using short sections of wire, which we will refer to as pixels, which are arranged in a 2D grid to create the detection area.  Such a design would use a large number of pixels thus giving high number resolution and the small size of the pixels gives short relaxation times.   We call this configuration a multi-layer superconducting number-resolving photon detector.

\section{A three layer architecture}
A significant problem with creating a two dimensional array of nanowire pixels is the question of how to attach the leads to each pixel.
One could simply move the pixels far enough apart to fit in all the necessary connections but this is impractical as the space between pixels would not detect photons and the device's overall efficiency would decrease below useful levels.  Ideally the pixels will be packed as closely as possible, while still avoiding cross talk, to maximize the so called fill factor.

\begin{figure}[htbp]
  \centerline{\includegraphics[width=0.6\textwidth]{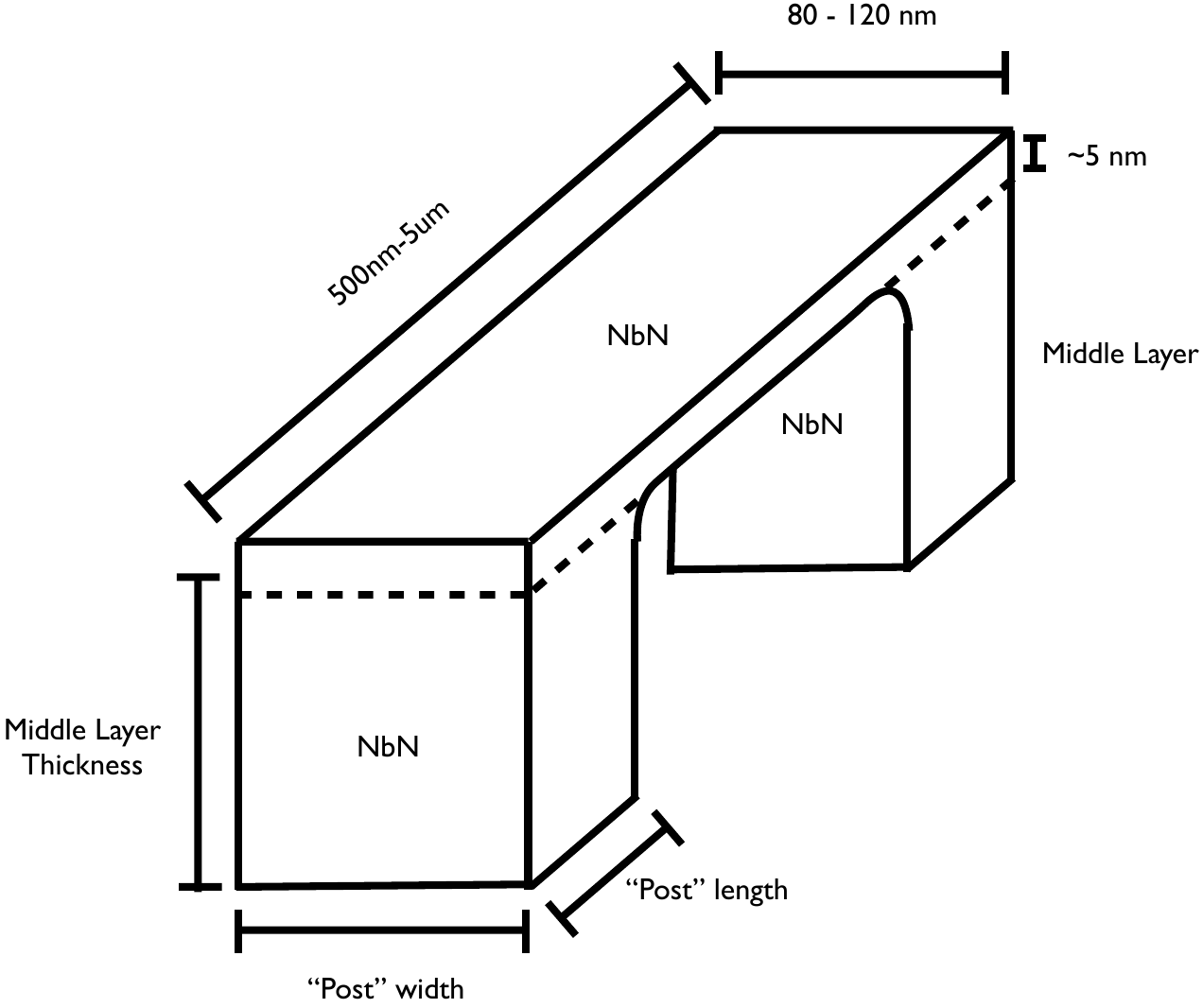}}
   \caption{{\bf A Single Pixel.}   One superconducting nanowire bridge forms a single pixel, NbN in this example.  The dimensions of the nanowire are fairly standard, 80 to 120 nm wide, 5nm thick and approximately several hundred nanometers long.  The sections of the nanowire in the middle layer are referred to herein as ``posts". Not to scale.}
\label{bridge}
\end{figure}

  We therefore propose moving away from the two dimensional approaches used to date and instead suggest a multi-layer, three dimensional architecture.  In this scheme the non-superconducting leads are allowed to pass under the active detector pixels.  To create this effect we shape the pixels like small bridges as seen in figure (\ref{bridge}).  This shape was chosen because of its relatively simple design and in order to maximize the fill factor.  The width and thickness of the wire is no different from that of a normal nanowire detector.  The minimum length of the bridge will have to be determined by the number of leads that are designed to pass underneath each pixel and need not be constant.

\begin{figure}[htbp]
   \centerline{\includegraphics[width=0.5\textwidth]{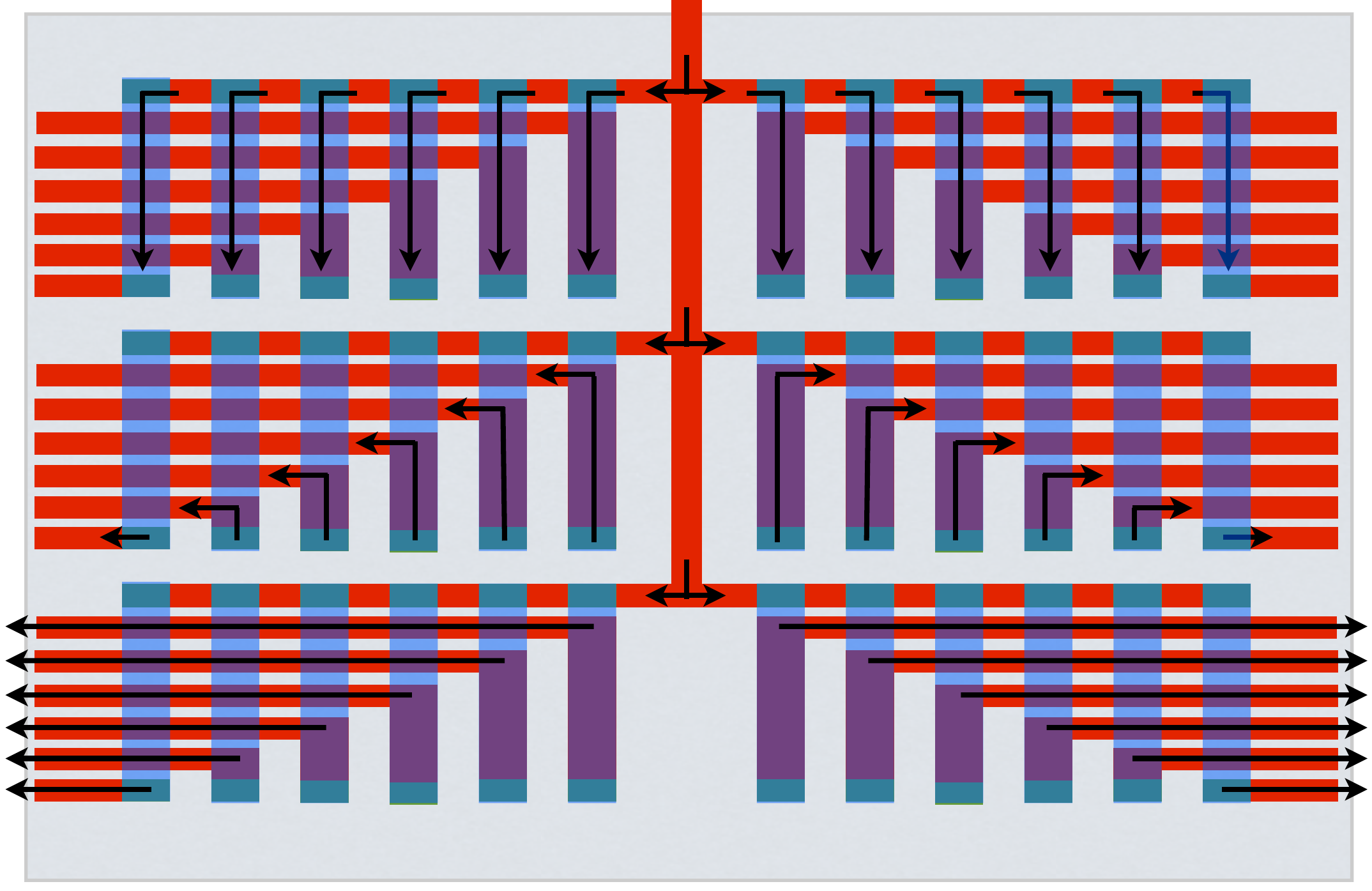}}
   \caption{{\bf Pixel Detector.} (Color Online) An plan view of a multilayer superconducting pixel detector.  Purple bars are the top layer superconducting NbN nanowire pixels.  Red wires are non superconducting leads that pass under the detection layer.  Green are the vertical connections between the bottom layer and the detection layer.  (Black) Arrows indicate the direction of the bias current flow throughout the device.}
\label{pd}
\end{figure}

  
This creates a 3 layer design, with the bottom layer containing the leads, an insulating middle layer and the active detection layer on top.  We now show a birds eye view (plan view) of the final device with all three layers aligned on top of each other, figure (\ref{pd}).  The black arrows show the movement of the current throughout the device. The current enters the detector in the bottom layer through the shared input lead (center of figure).  As an aside each pixel can be wired as an independent circuit, but the number of leads will increase and counting simultaneous events between elements becomes difficult.  The current then moves up through the middle layer connections, called ``post",  to the top/detection layer.  Once in the top layer it moves along the surface of the bridge.  This is the area in which an incident photon will form a resistance blockage.  Then it moves back down to the bottom layer and is channeled out of the device by the output leads.  Note that the red leads pass under the detectors, between the posts and that the input/output leads are all on the same layer.  The external detection electronics would then be similar to that proposed by Divochiy {\it et al.} \cite{divochiy}.

To determine the detection area we provide an example of relatively standard dimensions for NbN nanowire detectors \cite{gurevich,dauler,kerman,aja,aja2,kerman2}.  The device shown in figure (\ref{pd}) has six pixels in each arm and 6 arms.  If each wire is 100 nm wide and the device has a fill factor of $50\%$ then each arm will be approximately 1100nm wide.  If each of the red leads in the bottom layer is also 100nm wide with 100nm spacings, then each arm is at least 1300nm long.  These dimensions mean the the in figure (\ref{pd}) has a minimum detection area of approximately $2.2\mu$m x $3.9\mu$m.  This area is highly adjustable.  To increase the width we add more pixels, which will also increase the length as more leads will be required in the bottom layer.  To increase the length one simply lengthens the pixels in the detection layer or adds additional arms.  This means the meander detection areas of interest, about $10\mu$mx$10\mu$m, are easily reproducible.

\begin{figure}[htbp]
   \centerline{\includegraphics[width=0.5\textwidth]{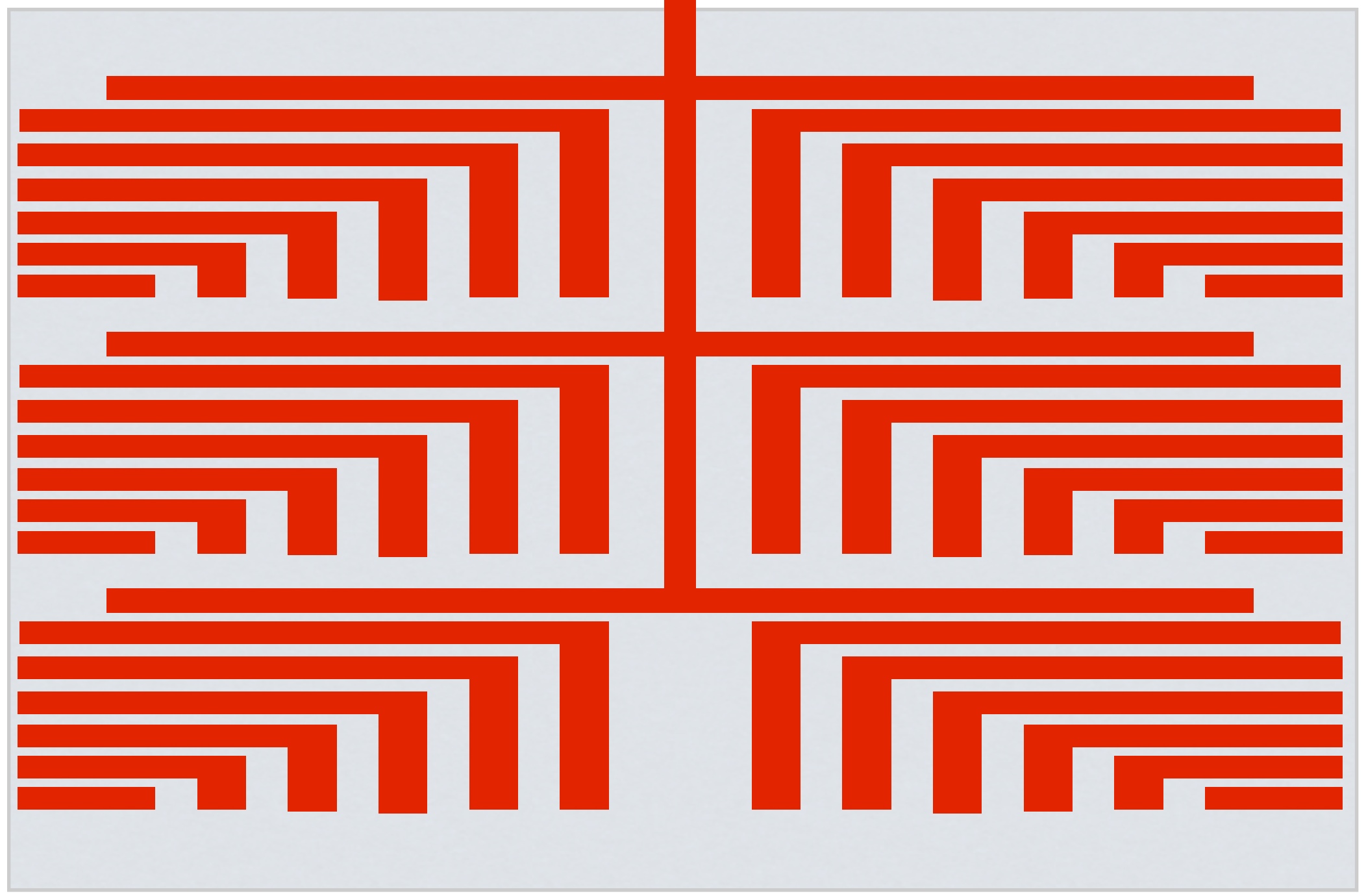}}
   \caption{{\bf Bottom Layer.} (Color Online) A plan view of the red non superconducting leads on a substrate that form the bottom layer and connect all of the pixels.  Note that there is no complete circuit on this layer.}
\label{bl}
\end{figure}

We now take a closer look at the minimum three layers needed to create the device.  The bottom layer, figure (\ref{bl}), consists of non superconducting leads placed on a insulating substrate, such as sapphire or MgO.  Note that there is no complete circuit on this level, so the current will be force to move up to the next level.  The optimal minimal spacing will depend on the insulating ability of the substrate to prevent leakage and cross talk, mainly between the input and output channels but also with the superconducting nanowires passing above.

\begin{figure}[htbp]
   \centerline{\includegraphics[width=0.5\textwidth]{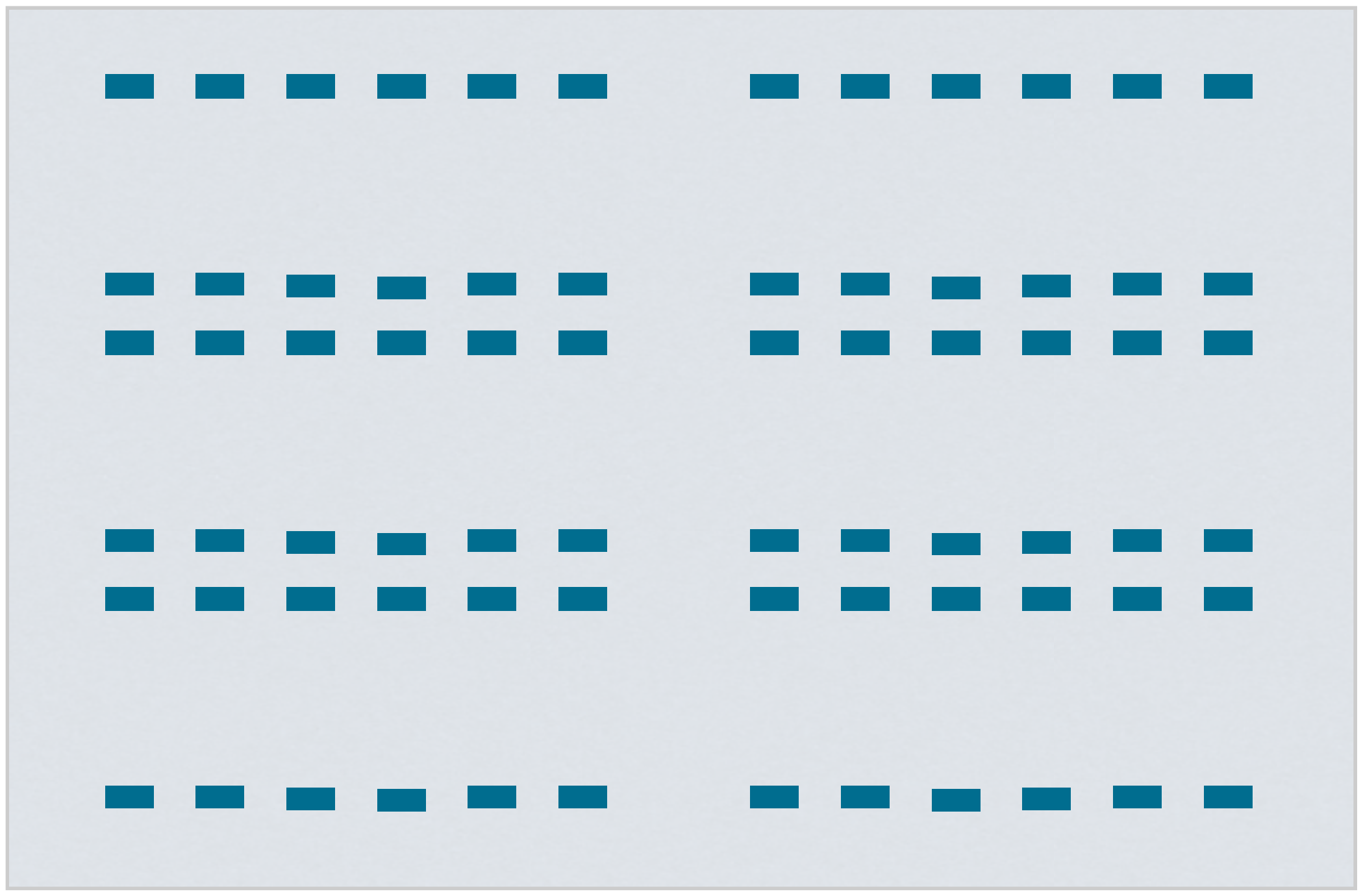}}
   \caption{{\bf Middle Layer.} (Color Online) A plan view of the vertical connections, ``posts", between the bottom layer and the detection layer built in to the substrate of the middle layer. these vertical section connect the leads to the nanowire detectors.}
\label{ml}
\end{figure}

Over the bottom layer will be a second layer of substrate shown in figure (\ref{ml}).  This layer will then have holes etched completely through it at predetermined locations so as to hit the input output leads in the bottom layer.  These holes are then filled with superconducting material (in practice it may be advantageous to us non-super conducting material here) thus completing the middle layer of the device.  Alignment will be very important but not an insurmountable issue as these structures are on the scale of approximately $100$nm in width and current alignment techniques can achieve results on the order of one nanometer or less\cite{anderson}.

Finally the detection layer will be deposited on top of the middle layer.  Alignment of the superconducting bridges with the vertical ``posts" in the middle layer will be important for the overall detection efficiency \cite{kerman3}.  See Fabrication below.

\begin{figure}[htbp]
   \centerline{\includegraphics[width=0.5\textwidth]{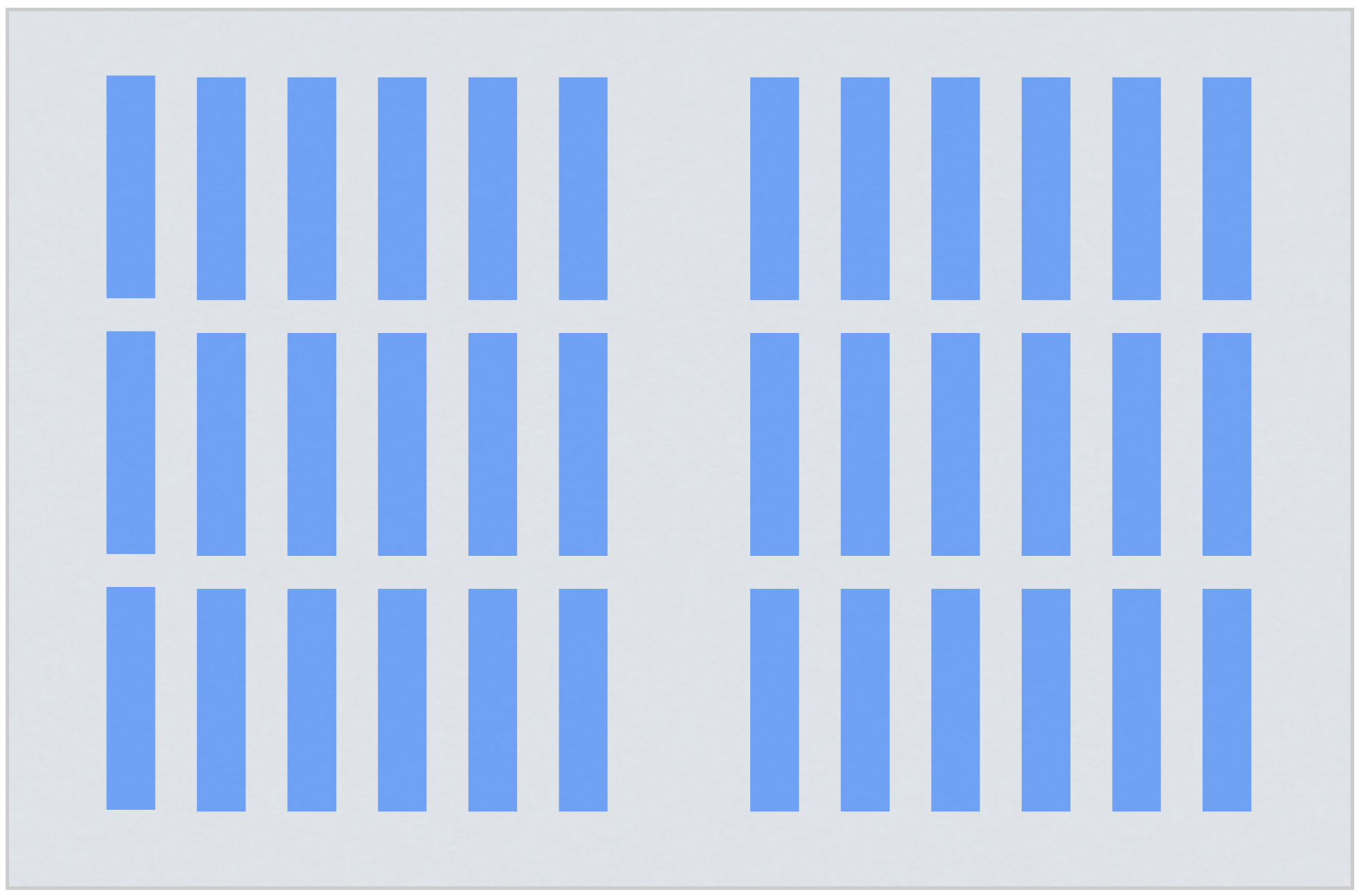}}
   \caption{{\bf Top Layer.} (Color Online) A plan view of the top layer of the detector.  This layer consists of the nanowires pixels that form the active area.  Note each pixel overlaps the posts in the middle layers.}
\label{tl}
\end{figure}

\section{Detector Characteristics}
  The characteristics of nanowire detectors have been studied in depth elsewhere \cite{gurevich,dauler,marsili,kerman,aja,aja2,kerman2,clem,kerman3}.  We will not repeat this fundamental work but will instead characterize the unique qualities of this device.  We will first look at number resolving power of our device vs the standard single or multi nanowire meander, as the purpose of this design is to increase the resolution.

From Dauler {\it et al.} we see that the probably of successfully counting $n$ out of $n$ photons by a detector with $D$ identical elements is simply \cite{dauler},

\begin{equation}
\label{mats}
P(n|n)=\frac{D!}{D^n(D-n)!}\eta^n.
\end{equation}

Here $\eta$ is the overall or average detection efficiency of each element.  The first half of this term is simply the probability that $n$ identical balls (photons) will fall into $D$ identical boxes (pixels) such that there is never more than 1 ball in any given box or $D$ choose $n$. Clearly increasing the number of detectors $D$ will significantly increase the number resolving power.  Also the most photons any such device can ever hope to resolve is equal to the number of detectors.  In their paper Dauler {\it et al.} created and tested a 4 wire meander that covered about $10\mu$m x $10\mu$m.  Consequently they have the following probabilities of correctly detecting 1,2,3,4 or more photons, given the assumptions of a 100\% fill factor and $\eta=100$\% detection efficiency.

\begin{table}[h]
 \caption{Successful Detection Probabilities.}
 \centering
 \begin{indented}
 \item[] \begin{tabular}{c c c} 
 \br
 Incident Photons, & 4 Wire Meander, & 36 Pixel Detector\\ [0.5ex]	
 \hline 
 1 & 1.0 & 1.0 \\ 
 2 & 0.75 & 0.972  \\ 
 3 & 0.375 & 0.918 \\ 
 4 & 0.093 & 0.842 \\
 5 & 0.0 & 0.748 \\  
  \hline 
  \end{tabular} 
  \end{indented}
  \label{numb-resolv} 
  \end{table}

In comparison we show the successful measurement probabilities of the detector in figure (\ref{pd}), with the same assumptions.  The results in Table \ref{numb-resolv} should come as no surprise as the detector has 36 separate and identical pixels compared to only 4 in the meander built by Dauler \cite{dauler}.

The next important factor is the time resolution, or the relaxation time due to any photon event.  Kerman {\it et al.} found that a 4$\mu$m x 6$\mu$m meander NbN nanowire of width 100nm had a $90\%$ reset time on the order of 1-10ns.  They also showed that the reset time was determined mainly by the kinetic inductance of the wire.  This inductance behaves linearly with the length of the wire.  In fact the kinetic inductance of a wire can be tuned significantly by altering the geometry of the nanowire \cite{aja}.  Therefore if all other parameters remain the same the easiest way of reducing the detectors reset time is to reduce the length of the wire.  In nanowire meanders this means deceasing the detection area or increasing the number of nanowires in the meander and may not be feasible.  By creating an array of pixels we are already fulfilling this design paradigm. As a side note, the chance two sequential photons enter the device, with shorter separation than the rest time, and happen to strike the same pixel is $1/D$  For our design above that would be approximately $2.8\%$, where as a 4 nanowire device would have a $25\%$ chance of missing the photon.

\section{Fabrication}
    The key step in the fabrication will be the alignment and the interface geometry between the three layers.  The individual layers themselves consist of relatively simple shapes that should present little problem for fundamental fabrication.  
    
    One question about the design is the bends in the nanowire bridges at the joints between the middle and top layer as shown in figure (\ref{bridge}).  The current density will increase as it passes around these corners, in what is known as current crowding.  Current crowding in superconducting nanowires has been studied by Clem and Berggren \cite{clem} for several bend geometries.  They discuss an optimal $90^\circ$ bend, in  a 2D plane of thin film, with curved interior edges that show no current crowding.  This occurs when the radius of the interior curve is very large compared to the thickness of the nanowire.  In our case that thickness is likely to be only $\approx5$nm in the detection layer.   Thus one possible change to the shape of the bridge could be rounding the joint between the middle and detection, shown in figure (\ref{bridge}).  This would turn the bridges into elongated arches and reduce the current crowding at the interface \cite{clem}.  The pixel bridge will be less susceptible to current crowding than standard meanders as there are no in plan $90^\circ$ or worse $180^\circ$ turns and the current density in the post will be lower than in the bridge.  There for while there may be some current crowding  at the junction the current density in the top layer should remain closer to the maximum current density.  This translates in to high efficiency.   Another possible change could be to build the middle layer ``posts" out of the same non superconducting material as the leads, thus eliminating the bends entirely but reducing the wire length.   This would shorten the wires so that latching might be and issue but would all but eliminate the current crowding issue do to the lack of bend in the pixel.
      
     Alignment will also play a part at this interface.  Assume there is a small offset, that maintains the electrical connection, between the middle and detection layers.  This will effectively constrict the wire and increase the current density further, similar to current crowding.  To maintain the superconductivity the critical current will be reduced by these effects.  This reduces the current density in the rest of the unconstricted wire which has the effect of reducing the quantum efficiency $\eta$ \cite{clem,kerman3}.  Also the narrowing of the nanowire may also effect the kinetic inductance mentioned above and in \cite{aja}.  The quality of the manufacture will therefore have a large effect on the quantum efficiency.
     
    There is also a question as to how efficient the extreme ends of the bridges will be.  Similar to the argument above, the current destiny at these points will be lower as the current transfers from the middle layer to the detection layer due to current crowding.  This lower current density may lead to the case that the absorption of a single photon is less likely to create a resistance blockage.  Another possibility is that the geometry of the bridge will be such that the hot spot created by the absorption of a single photon may be too small the create the resistance blockage.  We point out that the area effected by such a defect is significantly smaller than the total area of any pixel and can be thought of as simply a slight decrease in the over all fill factor.

\subsection{Pixel Size Limit}
   There is a physical limit on the minimum length of a nanowire detector.  This is based on the effect of ``latching" in superconducting nanowires.  If the return time ratio, $\tau_r=L_k/R_l$, of the kinetic inductance to the external resistance load becomes too small the nanowire can fail to return to is superconducting state after adsorbing a photon.  For specific details of latching in various systems see \cite{gurevich,aja2,kerman2}.  The easiest way to reduce the reset time $\tau_r$, which we take advantage of, is to simply make the nanowire shorter.  However due to latching there is a minimum length for any given wire geometry, width, thickness and external resistance load.  Altering and particularly lowering the external resistance load has other negative consequences that can out weigh the benefit of faster reset times.
   
     We point out that this is not a prohibitive limit to the construction of this device.  While we will have to be careful to avoid the latching condition we note that length of the superconducting nanowire in our pixel is not equal to the length of detection area.  The active detection area of the pixel on the detection layer is only part of the full length of the nanowire.
       There are two relatively free parameters in the design of the pixel bridge shown in figure (\ref{bridge}), both related to the design of the middle layer figure (\ref{ml}).  First is the width and length of the vertical ``posts" that connect the bottom layer to the detection layer.  We assume that the width of the ``post" will be the same as the width of the wire on the detection layer but the length of the post is then a relatively free parameter, with the constraint that the leads in the bottom layer must have room to pass under the bridge.  The second relatively free parameter is the thickness of the middle layer and therefore the height of the post.  This must be must be sufficient to prevent current leakage between the detection layer and the leads in the bottom layer that pass under them.  Increasing the thickness of the middle layer will increase the length of the wire and thus the reset time $\tau_r$ without effecting the detection area of any pixel or the fill factor of the array of pixels.  Annunziata {\it et al.} give a rule of thumb that 100nm wide NbN nanowires have a kinetic inductance of about 1nH/$\mu$m at 2.5K \cite{aja}.  Such a measurement can be used to customize the inductance and thus the reset time. The freedom in these two parameters, beyond their constraints, should give a measure of control over the reset time during the fabrication process.  This control may be enough to optimize the reset time while avoiding latching.  We also point out that there is a  freedom in selecting the superconducting material used in the detector.  It will be easier to avoid the latching condition by using a superconducting material that has a higher kinetic inductance per unit length than the current standard of NbN.  This is a different design paradigm than current meander devices which are trying to find low kinetic inductance materials in order to reduce the reset time.
     
\subsection{Pixel Number Limit}
  There is a practical limit to the number of pixels that can be placed in parallel for a given superconducting nanowire detector.  Assume that a device has a large number $D$ of nanowires in parallel, similar to figure (\ref{pd}). The bias current is set to be just under the critical current $I_c$ at which point the dc heating raises the temperature of the pixels out of the superconducting regime $I_c-I_b=\delta I$.  When a photon is incident on one detector element, the resistance in that element spikes and the current diverts into the external resistance load {\it and} into the other detector elements.  If a significant number of photons simultaneously trigger multiple elements the current increase in the other pixels may become larger than $\delta I$, which would cause them to erroneously detect $D$ photons \cite{divochiy}.

This effect has been taken advantage of in a different architecture to create Superconducting Nanowire Avalanche Photo-detectors (SNAPs), to increase the signal to noise ratio \cite{marsili}.
This effectively puts a limit on the number of pixels in parallel in a detector.  The maximum number of photons the device is capable of detecting may be less than the total number of detector elements.  However from table \ref{numb-resolv}, we see that the device would not be particularly efficient at detecting large numbers of photons as compared to the number of pixels anyway.  Also this failure state is easy to recognize should it occur and it is therefore easy to remove by post-selection.  We point out that very few applications require high number resolution, mainly because there are no good high number resolution detectors available.  This effect will be highly dependent on the external electronics of the detection system and can be mitigated by breaking up the detector area into several similar circuits.
Finally if the detector is being fabricated for a specific experiment the maximum number of photons should be well known and the number of pixels can be set to have a reasonable probability of successfully detecting all of them.


\subsection{The Spiral Meander}
\begin{figure}[htbp]
\centerline{\includegraphics[width=0.5\textwidth]{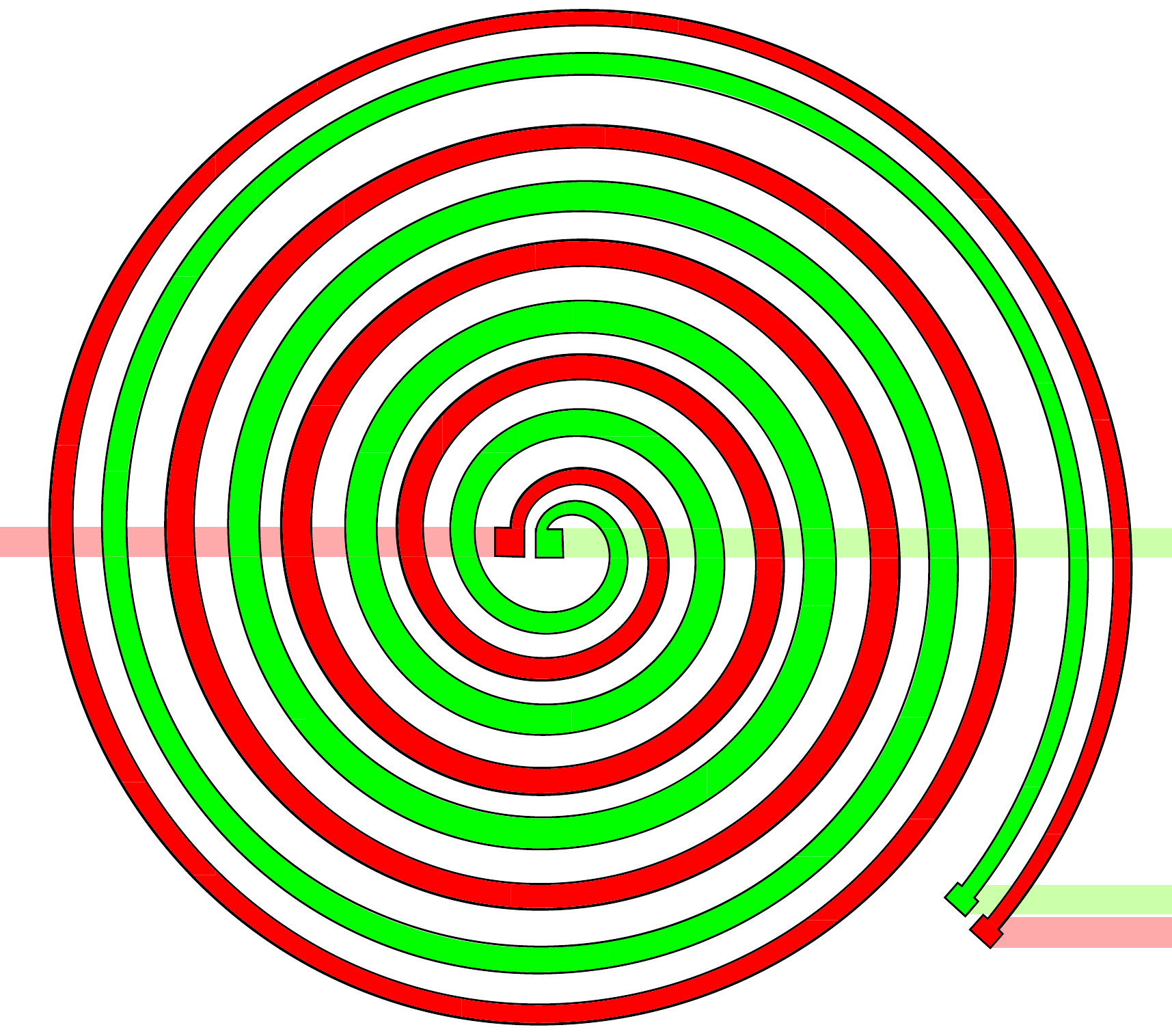}}
\caption{{\bf Top Layer.} (Color Online) An plan  view of a two element ``Spiral Meander" design.  The input/output (light red and green) leads will be connected via the bottom layer to the detection layer (dark red and green)}
\label{spiral}
\end{figure}

As another example of the versatility of the multi layer architecture we will briefly mention another 3D design noting that the shape of the pixels need not be limited to straight lines.  Another possible design using our multi-layer architecture is the Spiral Meander.  The Superconducting nanowires form concentric spirals  to create the detection area as sean in figure (\ref{spiral}).  In standard 2D architecture such a design would be impossible as there would be no way to connect the output leads to the center of the spiral.  We take advantage of our 3D architecture and create sinks in the center.  Similar to the vertical post in the bridge design, the sinks penetrate the middle layer and allow for the output leads to be connected from underneath the detection area.
  
Such a design will have trade-offs between it and the pixel array, or any other design.  For instance connecting the leads in the spiral meander will be easier then then large number of closely packed leads in the pixel array.  The spiral meander however does have fewer elements as shown here and therefore less number resolving ability than the pixel array.  This can be offset by putting more concentric wires in the spiral meander or by segmenting each path such that each wire in the detection layer covers a set solid angle or has set length and is then replaced in the pattern with a new wire, i.e. a new curved pixel.  This will also have the effect of reducing the wire length thus decrease the reset time.

The Spiral meander does have several potential drawbacks.   First the radius of curvature can not become infinitely small at the center.  Setting aside fabrication issues, there is a limit set by current crowding that will reduce the overall efficiency of the wire due to sharp curves \cite{clem}.  This means there is a finite sized hole in the center of the Spiral Meander, which we use to place the sinks.  This is could be a significant source of inefficiency as most inputs, such as single mode fibers, will have a gaussian like distributions centered at the center of the detection area.  This means a large number of photons may be lost depending on the hole size, the detection efficiency of the sinks, the spot size and the spot intensity distribution.
  
\section{Conclusion}
 If the fabrication issues mentioned above among others can be overcome, the multilayer superconducting number-resolving photon detector represents a significant improvement on current single layer meander devices.  The device will have significantly higher number resolution, even for large numbers of photons, while maintaing a useful detection area.  It has several parameters which can control the reset time to avoid latching while still minimizing the rest time.  An array of pixels of arbitrary size and shape is possible.  As mentioned above most of the detector will remain active after a single photon absorption as apposed to small number or single meander detectors which are effectively blinded by a single photon.  The active area of the detector can be tuned by changing the number or dimensions and or the shapes of the pixels and the fill factor of the detector should be at least equal to that of current nanowire meanders.  As a final note we will point out that the multi-layer superconducting number-resolving photon detector can also give a rough spatial distribution of the incident photons.  These advantages are compelling theoretical evidence for the construction and testing of multi-layer superconducting number-resolving photon detector.

\section*{Acknowledgments}
This work was supported by the National Research Council Research Associateship program at the AFRL Rome Research Site, Information Directorate.  We are grateful for useful discussions with C. Peters and M. Fanto among others.

Correspondence and requests for materials should be addressed to A.M.S. at email: amos.matthew.smith.ctr@rl.af.mil .

~

\end{document}